\documentclass
[a4paper,preprint,showkeys,showpacs,titlepage,NJP,final,onecolumn,endfloats,doublelines,12pt]{revtex4}
\usepackage{amsfonts}
\usepackage{amsmath}
\usepackage{amssymb}
\usepackage{graphicx}%
\providecommand{\U}[1]{\protect\rule{.1in}{.1in}}
\newtheorem{theorem}{Theorem}
\newtheorem{acknowledgement}[theorem]{Acknowledgement}

\begin{document}
\title[ ]{Time evolution of spin-boson system for different effective spectral density functions}
\author{Xian-Ting Liang}
\affiliation{Department of Physics and Institute of Modern
Physics, Ningbo University, Ningbo 315211, China}

\begin{abstract}
In this paper we firstly obtain two kinds of effective spectral
density functions by setting the cut-off frequencies of baths be
infinite and finite. Secondly, we investigate the reduced dynamics
of open qubits in four kinds of systems constructed with the basic
spin-boson model. It is shown that the qubit has different
dynamics governed by the two kinds of spectral density functions.
In addition, we obtained that a qubit coupled to an intermediate
harmonic oscillator has longer decoherence and relaxation times as
they are coupled to a common bath than to their respective baths.
In solving the dynamics of qubits we use a numerically exact
algorithm, iterative tensor multiplication algorithm based on the
quasiadiabatic propagator path integral scheme.
\end{abstract}
\startpage{1}
\endpage{102}
\maketitle

\section{Introduction}

As a quantum system of interest interacts with its environment
which is also considered being made up of a large number of the
quantum particles, solving the dynamics of the total system is in
fact an impossible task by now. However, in many cases we are only
interested in the dynamics of the quantum system of interest, thus
we can investigate the reduced system by using the
phenomenological environment as proposed by Caldeira and Leggett
\cite{AnnPhys149_374_1983}. Usually, in this method the
environment is modelled by a bath of harmonic oscillators (or
other quantum particles) and the influence of the bath on the
quantum system is fully enclosed in the so-called spectral density
function $J(\omega).$ According to the characteristic of the
environment one can choose the $J(\omega)$ Ohmic, sub-Ohmic,
super-Ohmic forms and so on. Thus, the reduced dynamics of the
quantum systems can be investigated in detail. Many such problems
have been treated extensively and some significative results have
been obtained in the field in past years
\cite{Weiss-book,RMP59_1_1987}.

The spectral density functions $J(\omega)$ obtained from the
microcosmic dynamical model plus its phenomenological environment
(MDMPPE) is in good agreement with that obtained from
quasi-classical equations of motion satisfied by the system of
interest. We can obtain the effective spectral density function
(ESDF) following Leggett as follows. Setting
$\sum_{i}c_{i}^{2}\delta\left( \omega-\omega_{i}\right)
/m_{i}\omega_{i}=J_{0}(\omega),$ taking cut-off frequency of the
bath modes infinity, and supposing the spectral density function
$J(\omega)$ a continuous function, we can obtain the ESDF with
$J(\omega)=$Im$[K(\omega)]$ (see Ref.\cite{PRB30_1208_1984}),
where $K(\omega)$ represents the effects of arbitrary linear
dissipative and/or reactive elements. Usually, $K(\omega)$ can be
obtained from the equation of motion of the system in
Fourier-transformed form, namely, $K(\omega)q\left(\omega\right)
=-\left[\partial V\left(q\right)/\partial
q\right]\left(\omega\right),$ where $V\left(q\right)$ is a
conservative potential. This equation can be obtained from the
scheme of MDMPPE. The ESDFs from infinite cut-off frequency of
baths for many models are obtained
\cite{JCP83_4491_1985,PhysRep168_115_1988,CPL449_296_2007}. This
method is in fact a universal one.

However, when the temperature of the systems are low enough, we
can imagine that the dynamics of the quantum systems of interest
obtained from the MDMPPE should have some difference with that
from quasi-classical equations of motions. In this case, the
high-frequency modes of the bath are in fact suppressed, so the
cut-off frequency may be not very large and it may be comparable
to the characteristic frequency of the quantum system of interest.
So there are at least two differences between the dynamics of
normal and low temperature for the quantum systems of interest.
The first is that the ESDFs obtained from the MDMPPE at low
temperature should not be in accordance with the spectral density
functions from quasi-classical equations of motion, and the second
is that the system at low temperature must go through a
non-Markovian process \cite{PRB78_235311_2008,JCP129_224106_2008}
other than the Markovian one.

With the progress of technique many quantum systems have been
designed and manufactured. In particular, many qubit models for
making quantum computers in future are proposed in recent years.
Most of these quantum systems should work in very low temperature
\cite{PRB75_104516_2007,PRB76_155323_2007}. The non-Markovian
effects of baths in the systems have been considered recently
\cite{PRA80_042112_2009,PRE80_041106_2009,PRA80_031143_2009,PRL103_050403_2009}.
It is reported that in many chemical and biologic systems, the
non-Markovian effects may also be important
\cite{ChemRev109_2350_2009,CP347_185_2008,Science316_1462_2007,Nature446_782_2007}.
At very low temperature, the cut-off frequencies of the bath
should be finite. Taking a finite cut-off frequency of the bath
rather than the infinite one, we shall obtain another kind of the
spectral density function. In the following we denote the ESDFs
from infinite, and from finite cut-off frequencies of the bath for
model $x$ by $J_{x}^{I}(\omega)$ (from infinite cut-off
frequencies) and $J_{x}^{F}(\omega)$ (from finite cut-off
frequencies), where $x$ denotes the physical models. A somewhat
similar discussion on the problem in completely classical
framework has been given in Refs
\cite{EPJB61_271_2008,PRE79_031128_2009}.

In this paper, we shall investigate the ESDFs of four models by
using the scheme of MDMPPE at infinite and finite cut-off
frequencies of baths. Through the two sets of ESDFs
$J_{x}^{I}\left(\omega\right)\ $ and
$J_{x}^{F}\left(\omega\right)$ we shall investigate the reduced
dynamics of qubits in some practical physical models. They are a
quantum system in a bath (model A); a quantum system coupled to an
intermediate harmonic oscillator (IHO) and only the latter coupled
to a bath (model B); a quantum system coupled to an IHO and both
of them coupled to their independent baths (model C) and a common
bath (model D).

\section{Two kinds of spectral density functions and the dynamics of open qubits}
In this section, we shall obtain two kinds of ESDFs for the models
A, B, C, and D when they are reduced to the original spin-boson
form. One kind of them are obtained from taking finite cut-off
frequencies of baths and the other is gotten by setting the
cut-off frequencies of baths infinity. After obtaining the
spectral density functions we shall solve the reduced dynamics for
the quantum systems of interest when they are set as the qubits.
The reduced dynamics governed by the two kinds of ESDFs for the
qubits will be compared.

In order to compare the dynamics of the qubit governed by the
different effective spectral functions in each model, we choose a
same starting point $J_{0}\left( \omega\right)$ for all of the
models. In another word, we always reduce them to the original
spin-boson model with the same IHO transformation and set
$J_{0}(\omega)=\sum_{i}c_{i}^{2}\delta\left(
\omega-\omega_{i}\right) /m_{i}\omega_{i}=\eta\omega
e^{-\omega/\omega_{c}}$ in our derivations. Because the model A is
just the original spin-boson model, our work for reducing this
model is just reducing it to itself. So in the assumption of
infinite cut-off frequency, the obtained ESDF of the model should
be equal to the $J_{0}\left(\omega\right)$, which will be shown in
the following subsection A. Models B, C, and D are different from
the original spin-boson model, so when they are reduced to the
original spin-boson form, the ESDFs should be different from the
starting function $J_{0}\left( \omega\right)$.

We have calculated the dynamics of the quantum system of interest
governed by $J_{x}^{I}\left(\omega\right)$ and
$J_{x}^{F}\left(\omega\right)$ with the same values of $\eta$ in
each models of the paper, and they are certainly different.
However, even for the same model, the different ESDFs
$J_{x}^{I}\left(\omega\right)$ and $J_{x}^{F}\left(\omega\right)$,
may provide different damping for the same $\eta$ in the function
$J_{0}(\omega)$, which will affect the dynamics of the quantum
system of interest. Do the different kinds ESDFs affect the
dynamics of quantum system of interest mainly through resulting in
different damping or through their different forms? To answer this
problem, we introduce an effective coupling strength
$\eta'=J(\omega=\omega_{0})/\omega_{0}$ to calculate the dynamics
as Ref.\cite{JCP96_8485_1992}. The effective coupling constant is
introduced from the physical deduce that the modes which has the
same frequency as the characteristic frequency of the system
$\omega_{0}$ has the major contribution to the system-bath
interaction. Thus, by using the effective coupling constant we can
obtain that, if the reduced dynamics governed by
$J_{x}^{I}\left(\omega\right)$ and $J_{x}^{F}\left(\omega\right)$
are different, we can safely attribute the different forms of the
$J_{x}^{I}\left(\omega\right)$ and $J_{x}^{F}\left(\omega\right)$.
For lack of space, in the following, we only plot the evolutions
of elements of reduced density matrix in the same effective
coupling strength $\eta'$ for each models.

\subsection{Two basic models (models A and B)}
A quantum system embeds in a bath (model A) and a quantum system
interacts with an IHO which is coupled to a bath (model B), are
two basic models and many complicated models are developed from
them. So in this subsection, we use them to show the Leggett's
method for obtaining the ESDFs $J_{A/B}^{I}\left( \omega\right)$,
and $J_{A/B}^{F}\left( \omega\right)$. Then from the two sets of
ESDFs we investigate the reduced dynamics of qubits in the two
models.

Model A: Because the physical quantities of the quantum system of
interest is not included in the ESDF, we can deduce the ESDFs
of the model A with the Hamiltonian as \cite{JCP83_4491_1985}%

\begin{equation}
H_{A}=\frac{p_{q}^{2}}{2\mu}+U\left(  q\right)  +%
{\textstyle\sum_{i}}
\frac{p_{i}^{2}}{2m_{i}}+\frac{1}{2}m_{i}\omega_{i}^{2}\left(  x_{i}%
+\frac{c_{i}q}{m_{i}\omega_{i}^{2}}\right)  ^{2}, \label{1-1}%
\end{equation}
where $p_{q}$ is the momentum conjugate to coordinate $q$.
Defining $U^{\prime}=dU/dt$ and using the dots for the time
derivatives, we can obtain the classical equations of the
motion as%
\begin{align}
\mu\ddot{q}  &  =-U^{\prime}(q)-%
{\textstyle\sum_{i}}
c_{i}x_{i}-%
{\textstyle\sum_{i}}
\frac{c_{i}^{2}q}{m_{i}\omega_{i}^{2}},\label{1-2-1}\\
m_{i}\ddot{x}_{i}  &  =-m_{i}\omega_{i}^{2}x_{i}-c_{i}q. \label{1-2-2}%
\end{align}
Using the Fourier transforms, we can write Eqs.(\ref{1-2-1}) and (\ref{1-2-2}) as%
\begin{align}
-\mu\omega^{2}q\left(\omega\right)& =-U_{\omega}^{\prime}(q\left(
t\right)  )-%
{\textstyle\sum_{i}}
c_{i}x_{i}\left(  \omega\right)  -%
{\textstyle\sum_{i}}
\frac{c_{i}^{2}q\left(\omega\right)}{m_{i}\omega_{i}^{2}},\label{1-3-1}\\
-m_{i}\omega^{2}x_{i}\left(\omega\right) & =-m_{i}\omega_{i}^{2}%
x_{i}\left(\omega\right)-c_{i}q\left(\omega\right).\label{1-3-2}%
\end{align}
From Eq.(\ref{1-3-2}) we have%
\begin{equation}
x_{i}\left(\omega\right)=-\frac{c_{i}}{m_{i}\left(\omega_{i}^{2}%
-\omega^{2}\right)}q\left(\omega\right),\label{1-4}%
\end{equation}
Insetting Eq.(\ref{1-4}) into Eq.(\ref{1-3-1}), we have%
\begin{equation}
\left[-\mu\omega^{2}-\omega^{2}%
{\textstyle\sum_{i}}
\frac{c_{i}^{2}}{m_{i}\omega_{i}^{2}\left(  \omega_{i}^{2}-\omega^{2}\right)
}\right]q(\omega)=-U_{\omega}^{\prime}(q\left(  t\right)  ). \label{1-5}%
\end{equation}
Setting%
\begin{equation}
L\left(  \omega\right)  =%
{\textstyle\sum_{i}}
\frac{c_{i}^{2}}{m_{i}\omega_{i}^{2}\left(  \omega_{i}^{2}-\omega^{2}\right)
}, \label{1-6}%
\end{equation}
and
\begin{equation}
J_{0}\left(\omega^{\prime}\right)  =\pi\sum\frac{c_{i}^{2}}{m_{i}\omega_{i}%
}\delta\left(\omega^{\prime}-\omega_{i}\right)
\equiv\omega^{\prime}\eta
e^{-\omega^{\prime}/\omega c}, \label{1-7}%
\end{equation}
we then have%
\begin{align}
L\left(  \omega\right)   &
=\frac{1}{\pi}\int_{0}^{\infty}\frac{J_{0}\left(
\omega^{\prime}\right)  }{\omega^{\prime}\left(
\omega^{\prime2}-\omega
^{2}\right)  }d\omega^{\prime}\nonumber\\
&  =\frac{1}{\pi}\eta\int_{0}^{\infty}\frac{\exp(-\omega^{\prime}/\omega_{c}%
)}{\left(  \omega^{\prime2}-\omega^{2}\right)  }d\omega^{\prime}\equiv\frac
{1}{\pi}\eta R\left(  \omega\right)  , \label{1-8}%
\end{align}
where (see Appendix)%
\begin{equation}
R\left(\omega\right)=-\frac{\pi i}{\omega}e^{-\omega/\omega
c}-\frac{\pi
}{\omega}W\left(\omega\right). \label{1-9}%
\end{equation}
The expression of $ W\left(\omega\right)$ is given in the
Appendix. Substituting Eq.(\ref{1-9}) into Eq.(\ref{1-5}) we have%
\begin{align}
K\left(\omega\right)q\left(\omega\right) &  =\left[ -\mu\omega
^{2}+\omega\eta W\left(\omega\right)+\omega\eta ie^{-\omega/\omega
c}\right]q\left(\omega\right)\nonumber\\
&  =-U_{\omega}^{\prime}\left(q\right).
\end{align}
If taking $\omega_{c}$ be infinite, we can easily obtain the
spectral density function as
\begin{equation}
J_{A}^{I}\left(  \omega\right)=\omega\eta e^{-\omega/\omega c}.
\label{1-13}
\end{equation}
This is the conventional Ohmic spectral density function of bath
in the original spin-boson model and it has been widely used in
quantum dissipative systems. Here, we have only shown that,
starting from $J_{0}\left( \omega\right)$, an ESDF
$J_{A}^{I}\left( \omega\right)$ can be obtained. The $J_{0}\left(
\omega\right)$ and $J_{A}^{I}\left( \omega\right)$ are the same
for the bath in the original spin-boson model. The result shows
that the Leggett's method give a consistent result for the model.
If the model is not the original spin-boson model (as following
models B, C, and D), it can also be reduced to the original
spin-boson model but the ESDFs should not be equal to the
$J_{0}\left(\omega\right)$ even we start from the same
$J_{0}\left( \omega\right)$.

If taking $\omega_{c}$ be finite, we then obtain the ESDF as
\begin{equation}
J_{A}^{F}(\omega)=\operatorname{Im}(K(\omega))=\omega\eta\Theta\left(
\omega\right). \label{1-12}%
\end{equation}
Here, $\Theta\left(\omega\right)=\operatorname{Im}\left( W\left(
\omega\right)\right)  +e^{-\omega/\omega_{c}}.$ The other forms of
the spectral density functions, such as sub-Ohmic and super-Ohmic
forms can be obtained similarly. It shows that when the cut-off
frequency is finite, we should add the term
$\eta\omega\operatorname{Im}[W\left(\omega\right)]$ in the
spectral density function as Eq.(\ref{1-12}). It is interesting
that how does the added term influence the spectral density
function? In Fig.1 we plot
$\operatorname{Im}[W\left(\omega\right)]$ in different cut-off
frequencies of the bath. It shows that the absolute values of the
$\operatorname{Im}[W\left(\omega\right)]$ decreases with the
increase of $\omega_{c}$. When $\omega_{c}\rightarrow\infty $, the
$\operatorname{Im}[W\left( \omega\right)]\rightarrow0$. However,
when the cut-off frequency of the bath is not big enough, the
influence of $\operatorname{Im}[W\left(\omega\right)]$ should not
be neglected in the description of the environment.

In above derivation of the ESDFs, the physical quantities for the
quantum system of interest have not been involved, so the
$J\left(\omega\right)$ can control the dynamics of not only the
harmonic oscillator as in Eq.(\ref{1-1}) but also other quantum
systems. In order to check the differences of the influences that
above two different ESDFs $J_{A}^{I}\left(\omega\right)$ and
$J_{A}^{F}\left(\omega\right)$ impose on the dynamics for the
systems of interest, we firstly investigate a simple quantum
system, qubit. For the open qubit the total Harmiltonian reads
\begin{equation}
H=\frac{\hbar}{2}\left(  \epsilon\sigma_{z}+\Delta\sigma_{x}\right)  +%
{\textstyle\sum_{i}}
\frac{p_{i}^{2}}{2m_{i}}+\frac{1}{2}m_{i}\omega_{i}^{2}\left(  x_{i}%
+\frac{c_{i}\sigma_{z}}{m_{i}\omega_{i}^{2}}\right)  ^{2}. \label{1-14}%
\end{equation}
This is one of the simplest open models of quantum system but its
reduced dynamics can still not be solved exactly, and some
approximations must be appealed to. Many investigations on this
topic have been made in recent years
\cite{JCP126_114102_2007,JCP130_164518_2009,JCP130_204512_2009,JPSJ75_082001_2006,JPSJ78_073802_2009}.
As we stressed that when the cut-off frequency is not too large to
compare to the characteristic frequency of the qubit, the
non-Markovian effect is non-negligible. The well established
iterative tensor multiplication (ITM) algorithm based on the
quasiadiabatic propagator path integral (QUAPI) scheme is a good
tool for solving the reduced dynamics of quantum system of
interest. The ITM algorithm is a numerically exact one and is
successfully tested and adopted in various problems of open
quantum systems
\cite{JCP102_4600_1995,CPL221_482_1994,PNAS93_3926_1996,PRE62_5808_2000,PRB72_245328_2005}.
For details of the scheme, we refer readers to previous works
\cite{JCP102_4600_1995,CPL221_482_1994,PNAS93_3926_1996,PRE62_5808_2000,PRB72_245328_2005}.
To make the calculations converge we use the time step $\delta
t=0.1/\Delta$ which is shorter than the correlation time of the
bath and the characteristic time of the qubit. In order to include
as much non-Markovian effect of the bath as we can, and avoid the
heavy calculation load we choose $\delta k_{\max}=3$. Here,
$\delta k_{\max}\times \delta t$ is the memory times being
included in the calculations. In this paper we set $T=300$ k, and
the initial state of the bath be the thermal equilibrium state,
namely, $\rho _{bath}\left( 0\right) =e^{-\beta H_{b}}/$Tr$\left(
e^{-\beta H_{b}}\right)$. Here, $\beta=1/k_{B}T$, $k_{B}$ is the
Boltzman constant. As we calculate the evolutions of diagonal
elements of the reduced density matrix $\rho_{11}(t)$, we set
$\Delta=1\times10^{12}$ Hz, $\epsilon=0.01\Delta$, and the initial
state of the qubit be $\rho _{1}\left( 0\right) =\left\vert \Phi
_{0}\right\rangle \left\langle \Phi _{0}\right\vert $. As we
calculate the evolutions of non-diagonal elements of the reduced
density matrix $\rho_{12}(t)$, we set $\epsilon=1\times10^{12}$
Hz, $\Delta=0.01\epsilon$, and the initial state of the qubit be
$\rho _{2}\left( 0\right) =\left\vert \Psi _{0}\right\rangle
\left\langle \Psi _{0}\right\vert $. Here, $\left\vert
\Phi_{0}\right\rangle=\left\vert 0\right\rangle$ and $\left\vert
\Psi_{0}\right\rangle=1/\sqrt{2}\left( \left\vert 0\right\rangle
+\left\vert 1\right\rangle \right)$. The $\left\vert
0\right\rangle $ and $\left\vert 1\right\rangle$ are the basis
states of the qubit. The comparison pictures on $\rho_{11}(t)$ and
$\rho_{12}(t)$ in different ESDFs ($J_{A}^{I}(\omega)$, and
$J_{A}^{F}(\omega)$) and different cut-off frequencies of the bath
($\omega_{c}=4.0\Delta,4.1\Delta,4.3\Delta,$ and $10.0\Delta$) are
plotted in Fig.2, where $\eta'$ is set to be 0.004.

It is shown that in the model A  the qubit has shorter decoherence
and relaxation times governed by the $J_{A}^{I}(\omega)$ than by
$J_{A}^{F}(\omega)$.

Model B: There is another basic model that the quantum system of
interest is not coupled to any bath but coupled to an IHO which is
coupled to
a bath. The Hamiltonian of this model reads%
\begin{align}
H_{B}  &  =\frac{p_{q}^{2}}{2\mu}+U\left(  q\right)  +\frac{P^{2}}{2M}%
+\frac{1}{2}M\Omega_{0}^{2}\left(  X+\lambda q\right)  ^{2}\nonumber\\
&  +%
{\textstyle\sum_{i}}
\frac{p_{i}^{2}}{2m_{i}}+\frac{1}{2}m_{i}\omega_{i}^{2}\left(  x_{i}%
+\frac{\kappa_{1}c_{i}X}{m_{i}\omega_{i}^{2}}\right)  ^{2}. \label{2-1}%
\end{align}
In the model, the pure Hamiltonian of the quantum system of
interest is $p_{q}^{2}/2\mu+U\left(q\right)$. The IHO has mass $M$
and frequency $\Omega_{0}$, and is coupled by the quantum system
of interest with strength $\lambda$. The bath coupled to the IHO
is comprised of a set of harmonic oscillators, and their mass,
momentum, coordinate, and coupling coefficients are denoted by
$\{m_{i},p_{i},x_{i} ,c_{i}\}$. The introduced $\kappa_{1}$ is a
controlling parameter for discussions. It has been shown that the
model B can be reduced to the model A through an ESDF of the bath
\cite{JCP83_4491_1985, CPL449_296_2007}. Setting the cut-off
frequency of the bath be infinite, we can obtain the ESDF as
\begin{align}
J_{B}^{I}\left(\omega\right)&
=\frac{\lambda^{2}\Omega_{0}^{4}\kappa
_{1}^{2}\omega\eta}{\left(\omega^{2}-\Omega_{0}^{2}\right)^{2}%
e^{\omega/\omega_{c}}+\kappa_{1}^{2}\Gamma^{2}\omega^{2}e^{-\omega/\omega_{c}}%
},\label{2-2-2}
\end{align}
which has been obtained in Ref. \cite{CPL449_296_2007}. Here,
$\Gamma=\kappa_{1}\eta/M $. If the cut-off frequency is finite,
the ESDF becomes
\begin{align}
J_{B}^{F}\left(\omega\right)&
=\frac{\lambda^{2}\Omega_{0}^{4}\kappa
_{1}^{2}\omega\eta\Theta\left(\omega\right)}{\Xi\left(
\omega\right) ^{2}+\kappa_{1}^{2}\omega^{2}\Gamma^{2}\Theta\left(
\omega\right) ^{2}}.
\label{2-2-1}%
\end{align}
Here,
\begin{equation}
\Xi\left(\omega\right)
=-\omega^{2}+\Omega_{0}^{2}+\kappa_{1}\omega
\Gamma\operatorname{Re}[W\left(\omega\right)]. \label{2-3}%
\end{equation}
If the quantum system of interest is a qubit the model B can be
reduced to a spin-boson model as Eq.(\ref{1-14}) with ESDF
$J_{B}\left(\omega\right)$. By using the ITM algorithm we can also
investigate the reduced dynamics of the qubit. The evolutions of
the diagonal and off-diagonal elements of the reduced density
matrix of the qubit governed by $J_{B}^{I}\left(\omega\right)$ and
$J_{B}^{F}\left(\omega\right)$ are plotted in Fig.3 for the cases
of $\omega_{c}=3.0\Delta,5.0\Delta,10.0\Delta,$ and $25.0\Delta$.
Here, we set $\Omega_{0}=52.0\Delta$, and $\eta'=0.0035$, and
other parameters are the same as Fig.2. It is contrary to the case
of model A, here the qubit has longer decoherence and relaxation
times governed by the $J_{B}^{I}(\omega)$ than by
$J_{B}^{F}(\omega)$.

As the cut-off frequency of the bath is not very large, the
$J_{A}^{I}\left( \omega\right)$ results in shorter decoherence and
relaxation times for the qubit in the models A than the
$J_{A}^{F}\left( \omega\right)$ does. However, in model B, to
compare to the $J_{B}^{F}\left( \omega\right)$, the
$J_{B}^{I}\left( \omega\right)$ causes longer decoherence and
relaxation times of the qubit, as the cut-off frequency of the
bath is not very large. When the cut-off frequency of the bath
increases to 5 times (for the model A) and 25 times (for the model
B) of the characteristic frequency of the qubit, the
$J_{A/B}^{F}\left( \omega\right)$ and the $J_{A/B}^{I}\left(
\omega\right)$ will lead to almost the same reduced dynamics of
the qubit for the respective models.

\subsection{Two developed models (models C and D)}

In this section we investigate the other two models C and D. These
two models are developed from above two basic models A and B.

Model C: The Hamiltonian of the model C reads%
\begin{align}
H_{C}& =\frac{p_{q}^{2}}{2\mu}+U\left(q\right)+\frac{P^{2}}{2M}%
+\frac{1}{2}M\Omega_{0}^{2}\left(X+\lambda q\right)^{2}\nonumber\\
& +%
{\textstyle\sum_{i}}
\frac{p_{i}^{2}}{2m_{i}}+\frac{1}{2}m_{i}\omega_{i}^{2}\left(  x_{i}%
+\frac{\kappa_{1}c_{i}X}{m_{i}\omega_{i}^{2}}\right)  ^{2}\nonumber\\
&  +%
{\textstyle\sum_{j}}
\frac{p_{j}^{2}}{2m_{j}}+\frac{1}{2}m_{j}\omega_{j}^{2}\left(  x_{j}%
+\frac{\kappa_{2}c_{j}q}{m_{j}\omega_{j}^{2}}\right)  ^{2}. \label{3-1}%
\end{align}
Here, the quantum system of interest interacts with the IHO and
both of them are coupled to their independent baths. The bath
($B_{1}$) coupled to the IHO and the bath ($B_{2}$) coupled to the
quantum system of interest are considered to be constructed with
two sets of harmonic oscillators and those mass, momentums,
coordinates, and coupling coefficients are denoted by
$\{m_{\alpha},$ $p_{\alpha},$ $x_{\alpha},$ $c_{\alpha}\}$
($\alpha=i$ for $B_{1}$, and $\alpha=j$ for $B_{2}$). As
$\kappa_{1}$, the $\kappa_{2}$ is also a controlling parameter for
discussion. Similar to above subsection we can
obtain an ESDF for the model as%
\begin{align}
J_{C}^{I}\left(  \omega\right)   &
=\eta\omega\kappa_{2}^{2}e^{-\omega
/\omega_{c}}+\frac{\lambda^{2}\Omega_{0}^{4}\kappa_{1}^{2}\omega\eta}{\left(
\omega^{2}-\Omega_{0}^{2}\right)  ^{2}e^{\omega/\omega_{c}}+\kappa_{1}%
^{2}\Gamma^{2}\omega^{2}e^{-\omega/\omega_{c}}},\label{3-2-2}
\end{align}
when the cut-off frequency of the bath is infinite. Setting the
cut-off frequency of the bath be finite, we have the ESDF for the
model as
\begin{align}
J_{C}^{F}\left(\omega\right)&
=\eta\omega\kappa_{2}^{2}\Theta\left( \omega\right)
+\frac{\Omega_{0}^{4}\lambda^{2}\kappa_{1}^{2}\omega\eta
\Theta\left(\omega\right)}{\Xi\left(\omega\right)^{2}+\kappa_{1}%
^{2}\omega^{2}\Gamma^{2}\Theta\left(\omega\right)^{2}}. \label{3-2-1}%
\end{align}
If both $\lambda$ and $\kappa_{1}$ are equal to zero, namely the
quantum system of interest have not any interaction to other
systems except for being directly coupled to a bath (where the
bath is $B_{2}$) the model C will reduce to the model A, and from
Eqs.(21) and (22), we can obtain
\begin{align}
J_{C}^{I/F}\left(\omega\right)\xrightarrow{\lambda =\kappa _{1}=0}
J_{A}^{I/F}\left(\omega\right). \label{3-3-1}
\end{align}
When $\kappa_{2}=0$, the model C reduces to the model B which
describes a quantum system of interest interacting with its
environment through an IHO, and from Eqs.(21) and (22), we have
\begin{align}
J_{C}^{I/F}\left(\omega\right)\xrightarrow{\kappa _{2}=0}
J_{B}^{I/F}\left(\omega\right). \label{3-3-2}
\end{align}
Model D: Model C describes that the quantum system of interest
interacts with an IHO and both of them are coupled to two
independent baths. However, when the quantum system is very closed
to the IHO, they may embed in one common bath, then the
Hamiltonian of the total system becomes%

\begin{align}
H_{D}  &  =\frac{p_{q}^{2}}{2\mu}+U\left(  q\right)  +\frac{P^{2}}{2M}%
+\frac{1}{2}M\Omega_{0}^{2}\left(  X+\lambda q\right)  ^{2}\nonumber\\
&  +%
{\textstyle\sum_{i}}
\frac{p_{i}^{2}}{2m_{i}}+\frac{1}{2}m_{i}\omega_{i}^{2}\left[  x_{i}%
+\frac{c_{i}}{m_{i}\omega_{i}^{2}}\left(  \kappa_{1}X+\kappa_{2}q\right)
\right]  ^{2}. \label{4-1}%
\end{align}
Similarly, we can obtain an ESDF from infinite cut-off frequency
as
\begin{align}
J_{D}^{I}\left(\omega\right)& =\lambda M\Omega_{0}^{2}\operatorname{Im}%
\Phi+\omega\eta\left[
\kappa_{1}\kappa_{2}\operatorname{Re}\Phi+\kappa
_{2}^{2}\right]  e^{-\omega/\omega_{c}}.\label{4-4}%
\end{align}
Setting the cut-off frequency of the bath be finite, we have
\begin{align}
J_{D}^{F}\left(\omega\right)& =\lambda M\Omega_{0}^{2}\operatorname{Im}%
\Psi+\omega\eta\kappa_{1}\kappa_{2}\operatorname{Re}W\left(  \omega\right)
\operatorname{Im}\Psi\nonumber\\
&  +\omega\eta\left[\kappa_{1}\kappa_{2}\operatorname{Re}\Psi+\kappa_{2}%
^{2}\right]  \Theta\left(\omega\right), \label{4-2}%
\end{align}
where%
\begin{align}
\Phi & =\frac{\Omega_{0}^{2}\lambda+i\Gamma\kappa_{2}\omega
e^{-\omega/\omega_{c}}}{\left(  \omega^{2}-\Omega_{0}^{2}\right)
+i\kappa_{1}\omega\Gamma e^{-\omega/\omega_{c}}},\label{4-3-1}%
\end{align}
\begin{align}
\Psi & =\frac{\Omega_{0}^{2}\lambda+\Gamma\kappa_{2}\omega
\operatorname{Re}W\left(\omega\right)+i\Gamma\kappa_{2}%
\omega\Theta\left(\omega\right)}{-\Xi\left(\omega\right) +i\kappa
_{1}\omega\Gamma\Theta\left(\omega\right)}. \label{4-3-2}%
\end{align}
If we set $\kappa_{1}=0,$ $\lambda=0,$ and $\kappa_{2}=1,$ the
model D will reduce to the model A and from Eqs.(26) and (27), we
have
\begin{equation}
J_{D}^{I/F}\left(\omega\right)\xrightarrow{\lambda =\kappa
_{1}=0,\kappa_{2}=1}
J_{A}^{I/F}\left(\omega\right). \label{4-5}%
\end{equation}
Setting $\kappa _{2}=0,$ the model D will reduce to the model B
and from Eqs.(26) and (27), we can obtain
\begin{equation}
J_{D}^{I/F}\left(\omega\right)\xrightarrow{\kappa_{2}=0}
 J_{B}^{I/F}\left(\omega\right).\label{4-4}%
\end{equation}
By using the ESDFs $J_{C}^{I/F}\left(\omega\right)$ and
$J_{D}^{I/F}\left(\omega\right)$ we can compare the reduced
dynamics of the qubit in models C and D, as we have done in last
subsection. We can plot above different spectral density functions
as Fig.4. Here, we are more interested in the question: in which
model (C or D) the qubit has longer decoherence and relaxation
times when the values of all of the parameters in the models are
the same? Using the spectral density functions $J_{C}^{I/F}\left(
\omega\right)$ of the model C, and
$J_{D}^{I/F}\left(\omega\right)$ of the model D, we solve the
reduced dynamics of the qubit in the two models. The evolutions of
the elements of the reduced density matrix of the qubit are
plotted in Fig.5. Here, we set $\kappa_{1} =\kappa_{2}
=\lambda=1$, $\Omega_{0}=10\Delta$, and $\omega_{c}=7\Delta$.
Other parameters are set with the same values as in Fig.2. It is
shown that the qubit in the model D has longer decoherence and
relaxation times than it has in the model C at the same finite
cut-off frequency of the baths.

\section{Discussions and Conclusions}

The spectral density function is in fact a bridge to connect the
microcosmic and macroscopical physical quantities in dissipative
quantum theory. At low temperature, the excitations of the bath
modes with high frequencies are suppressed. So the spectral
density function derived from infinite cut-off frequency of the
bath will be out-of-true. In this paper we investigate the ESDFs
$J_{x}^{I/F}\left( \omega\right)$ for models ($x=$A, B, C, D) at
infinite and finite cut-off frequencies of baths from the
microcosmic dynamical models by using the method initiated by
Leggett et. al. As the cut-off frequency of the bath is comparable
to the characteristic frequency of the quantum system of interest,
the non-Markovian effects may also be important. In this case, it
is improper to use any Markovian approximations to investigate the
the reduced dynamics of quantum system. So we used a numerically
exact ITM algorithm based on the QUAPI scheme solving the reduced
dynamics of qubits.

In order to clearly compare the difference of the dynamics of
quantum systems resulting from different ESDFs $J_{x}^{I}\left(
\omega\right)$ and $J_{x}^{F}\left( \omega\right)$, we used the
effective coupling constants calculating the dynamics of the qubit
in each models and fixed other parameters such as the tempurature
is set $300$k. It is shown that a qubit (the quantum system of
interest) in these models has different dynamics when they are
governed by the spectral density functions $J_{x}^{I}\left(
\omega\right)$ and $J_{x}^{F}\left(\omega\right)$ as the cut-off
frequencies are not too large. The difference increase with the
decrease of the cut-off frequency. When the cut-off frequencies of
the baths are larger about 5 times (for the models based on model
A) and 25 times (for the models based on model B) than the
characteristic frequency of the qubit, the two kinds of spectral
density functions result in almost the same reduced dynamics. In
addition, we obtain that a qubit will have longer relaxation and
decoherence times when it is coupled to an IHO and both of them
are coupled to a common bath rather than to their respective
baths.

\begin{acknowledgement}
This project was sponsored by National Natural Science Foundation
of China (Grant No. 10675066), Natural Science Foundation of
Ningbo City (Grant No.2008A610098) and K C Wong Magna Foundation
in Ningbo University.
\end{acknowledgement}

\section{Appendix}

In the Appendix we give a proof of Eq.(\ref{1-9}), namely prove
equation $R\left(\omega\right)=-\frac{\pi
i}{\omega}e^{-\omega/\omega c}-\frac{\pi
}{\omega}W\left(\omega\right).$ Setting%
\begin{equation}
z=x+iy,\text{ }f\left(z\right) =\frac{\exp(-z/\omega_{c})}{\left(
z^{2}-\omega^{2}\right)},\label{a1}%
\end{equation}
and from Fig.5 we see that there is not any singular point in the
loop of $(L+\overline{M_{2}0}+\overline{0M_{1}}+C)$. So by using
the Cauchy's theorem we have%
\begin{equation}
\underset{G_{I}}{\underbrace{\int_{L}f(z)dz}}+\underset{G_{II}}{\underbrace
{\int_{R}^{0}f(iy)d\left(  iy\right)}}+\underset{G_{III}=R\left(
\omega\right)}{\underbrace{\int_{0}^{R}f(x)d\left(x\right)}}%
+\underset{G_{IV}}{\underbrace{\int_{C}f(z)dz}}=0. \label{a2}%
\end{equation}
For curve $L$, because%
\begin{equation}
\lambda_{1}=\lim_{R\rightarrow\infty}zf(z)=\lim_{R\rightarrow\infty}\frac
{\exp(-z/\omega_{c})}{\left(  z-\omega^{2}/z\right)}=0, \label{a3}%
\end{equation}
we have%
\begin{equation}
G_{I}=\int_{L}f(z)dz=i\left(  \theta_{2}-\theta_{1}\right)\lambda_{1}=0.
\label{a4}%
\end{equation}
For curve $C$, we replace pramaters $z$ and $\omega$ with $r$ and $\varphi$
and setting $z-\omega=re^{i\varphi}.$ Because $\varphi_{2}=\pi,$ $\varphi
_{1}=0,$ and%
\begin{equation}
\lambda_{2}=\lim_{r\rightarrow0}zf(z)=\frac{\exp(-\omega/\omega_{c})}{2\omega
}, \label{a5}%
\end{equation}
we have%
\begin{equation}
G_{IV}=\int_{c}f(z)dz=i\left(\varphi_{2}-\varphi_{1}\right)
\lambda
_{2}=\frac{i\pi\exp(-\omega/\omega_{c})}{2\omega}. \label{a6}%
\end{equation}
For linear $\overline{M_{2}0}$, the integral becomes
\begin{equation}
G_{II}=\int_{R}^{0}f(iy)d\left(  iy\right)  =-\frac{i}{\omega}\int_{R}%
^{0}\frac{\exp(-i\frac{y}{\omega}\cdot\frac{\omega}{\omega_{c}})}{\left(
\frac{y}{\omega}\right)  ^{2}+1}d\left(  \frac{y}{\omega}\right)  . \label{a7}%
\end{equation}
Setting $\frac{y}{\omega}=x,\frac{\omega}{\omega_{c}}=m,$ we have%
\begin{align}
G_{II}  &
=-\frac{i}{\omega}\int_{R}^{0}\frac{\exp(-ixm)}{x^{2}+1}d\left(
x\right) \nonumber\\
&  =-\frac{i}{\omega}\int_{R}^{0}\frac{\cos mx}{x^{2}+1}d\left(  x\right)
-\frac{1}{\omega}\int_{R}^{0}\frac{\sin mx}{x^{2}+1}d\left(  x\right)  .
\label{a8}%
\end{align}
Here,%
\begin{align}
&  \int_{R}^{0}\frac{\cos mx}{x^{2}+1}d\left(  x\right) \nonumber\\
&  =-\int_{0}^{R}\frac{\cos mx}{x^{2}+1}d\left(  x\right)  =-\frac{\pi}{2}%
\exp(-\omega/\omega_{c}),\label{a9}\\%
&  \int_{R}^{0}\frac{\sin mx}{x^{2}+1}d\left(  x\right) \nonumber\\
&  =-Shi\left(  m\right)  \cosh\left(  m\right)
+\operatorname{Ci}\left( -im\right)  \sinh\left(  m\right)
+\frac{\pi i}{2}\sinh\left(  m\right)
\nonumber\\
&  \equiv\pi W\left(  \omega\right). \label{a10}%
\end{align}
with%
\begin{align}
Shi(m)  &  =\int_{0}^{m}\frac{\sinh(t)}{t}dt,\label{a12}\\%
\operatorname{Ci}(m)  &  =\int_{0}^{\infty}e^{-t}t^{(m-1)}dt+\ln(m)+\int
_{0}^{m}\frac{\cos(t)-1}{t}dt. \label{a13}%
\end{align}
So we have%
\begin{equation}
G_{II}=\frac{\pi i}{2\omega}e^{-\omega/\omega c}+\frac{\pi}{\omega}W\left(
\omega\right). \label{a14}%
\end{equation}
From Eqs.(\ref{a2}), (\ref{a4}), (\ref{a6}), and (\ref{a14}) we
have the integral in linear $\overline{0M_{1}}$ as
\begin{align}
G_{III}  &  =\int_{0}^{R}f(x)d\left(x\right)\nonumber\\
&  =-G_{I}-G_{II}-G_{IV}\nonumber\\
&  =-\frac{\pi i}{2\omega}e^{-\omega/\omega c}-\frac{\pi i}{2\omega}%
e^{-\omega/\omega c}-\frac{\pi}{2\omega}W\left(  \omega\right) \nonumber\\
&  =-\frac{\pi i}{\omega}e^{-\omega/\omega c}-\frac{\pi}{\omega}W\left(
\omega\right) \nonumber\\
&  =R\left(\omega\right). \label{a15}%
\end{align}

\newpage

\begin{figure}[tbp]
\begin{center}
\scalebox{0.35}{\includegraphics{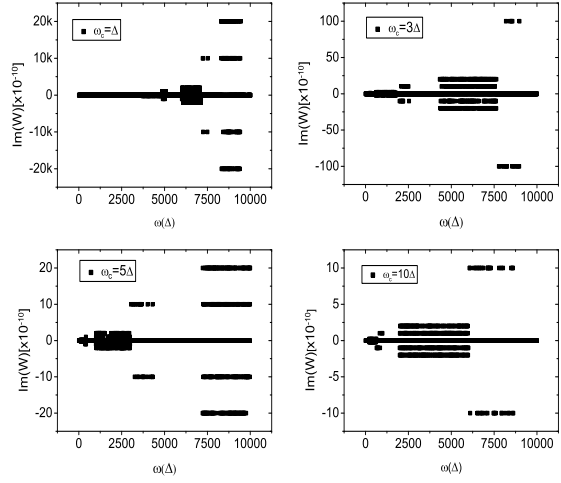}}
\end{center}
\caption{(Color online) The values of
$\operatorname{Im}[W\left(\omega\right)]$ when choosing the
cut-off frequency $\omega_{c}$ different values. It is shown that
the absolute values of $\operatorname{Im}[W\left(\omega\right)]$
decreases with the increase of the cut-off frequency of the bath
$\omega_{c}$}\label{fig1}
\end{figure}

\begin{figure}[tbp]
\begin{center}
\scalebox{0.3}{\includegraphics{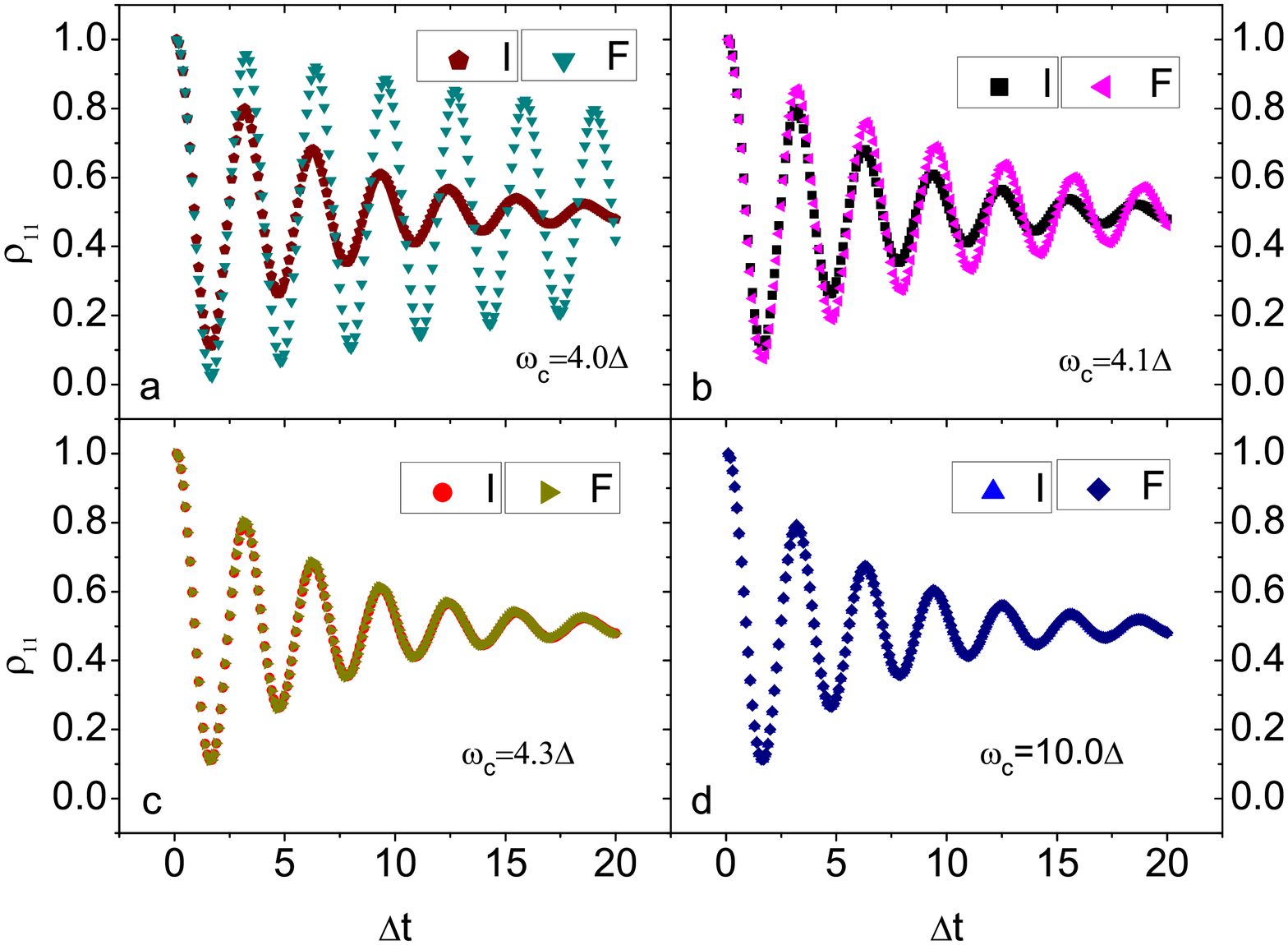}}
\scalebox{0.3}{\includegraphics{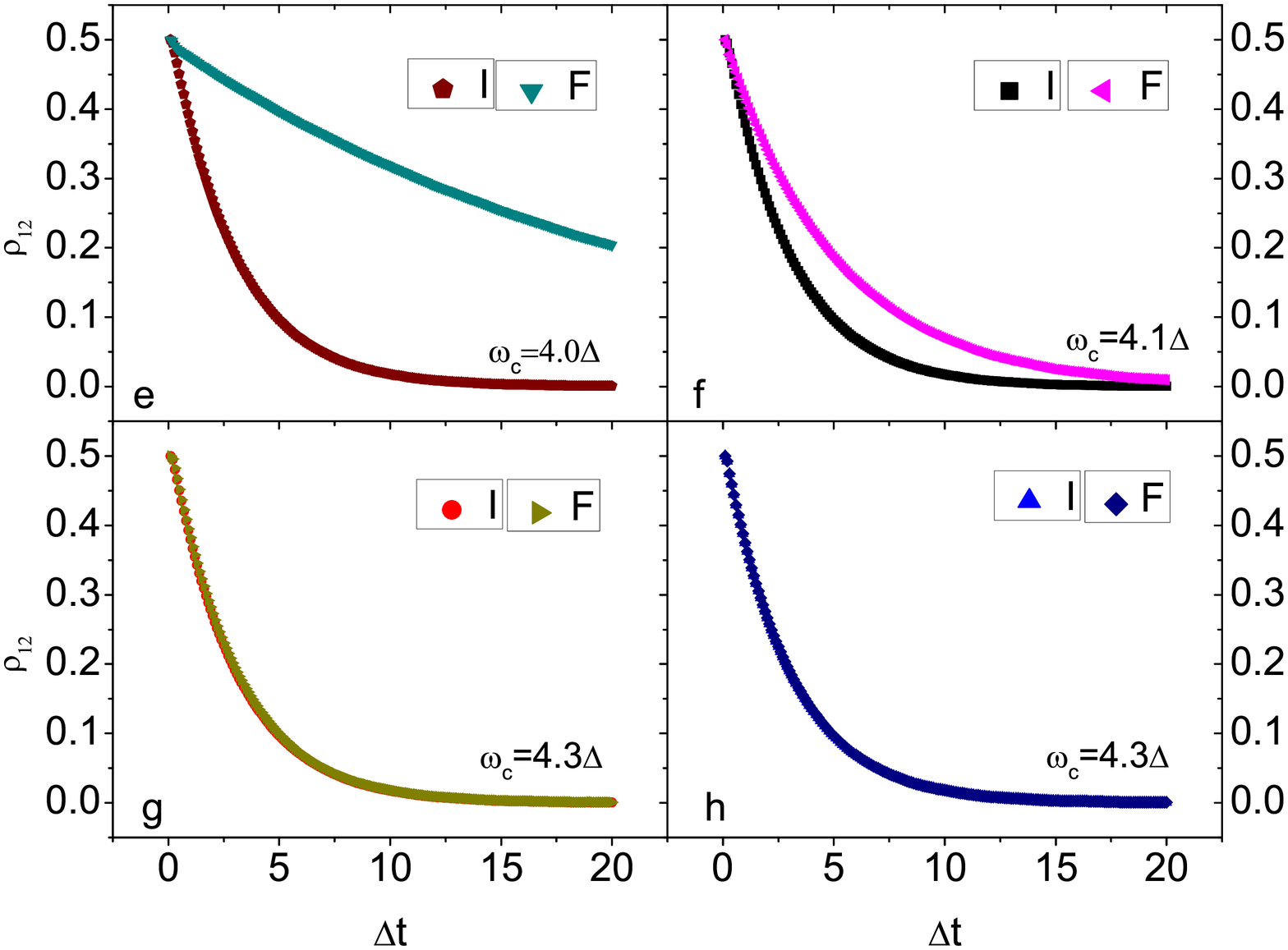}}
\end{center}
\caption{(Color online) The reduced dynamics of qubit governed by
$J_{A}^{I}\left( \omega\right)$ vs by
$J_{A}^{F}\left(\omega\right)$ in different values of
$\omega_{c}$. Top four figures a, b, c, and d plot the evolutions
of the diagonal element $\rho_{11}(t)$ beginning at the initial
state $\rho_{1}(0)$, where $\Delta=10^{12}$ Hz,
$\epsilon=0.01\Delta$. Bottom four figures e, f, g, and h plot the
evolutions of off-diagonal element $\rho_{12}(t)$ beginning at the
initial state $\rho_{2}(0)$, where $\epsilon=10^{12}$ Hz,
$\Delta=0.01\epsilon$, T=300 k, $\eta'=0.004$. The $\rho_{1}(0)$
and $\rho_{2}(0)$ are supposed in the text. Here, and in the
following, we denote the results from the ESDFs $J^{I}(\omega)$,
and $J^{F}(\omega)$ with I and F in the Legend} \label{fig2}
\end{figure}

\begin{figure}[tbp]
\begin{center}
\scalebox{0.3}{\includegraphics{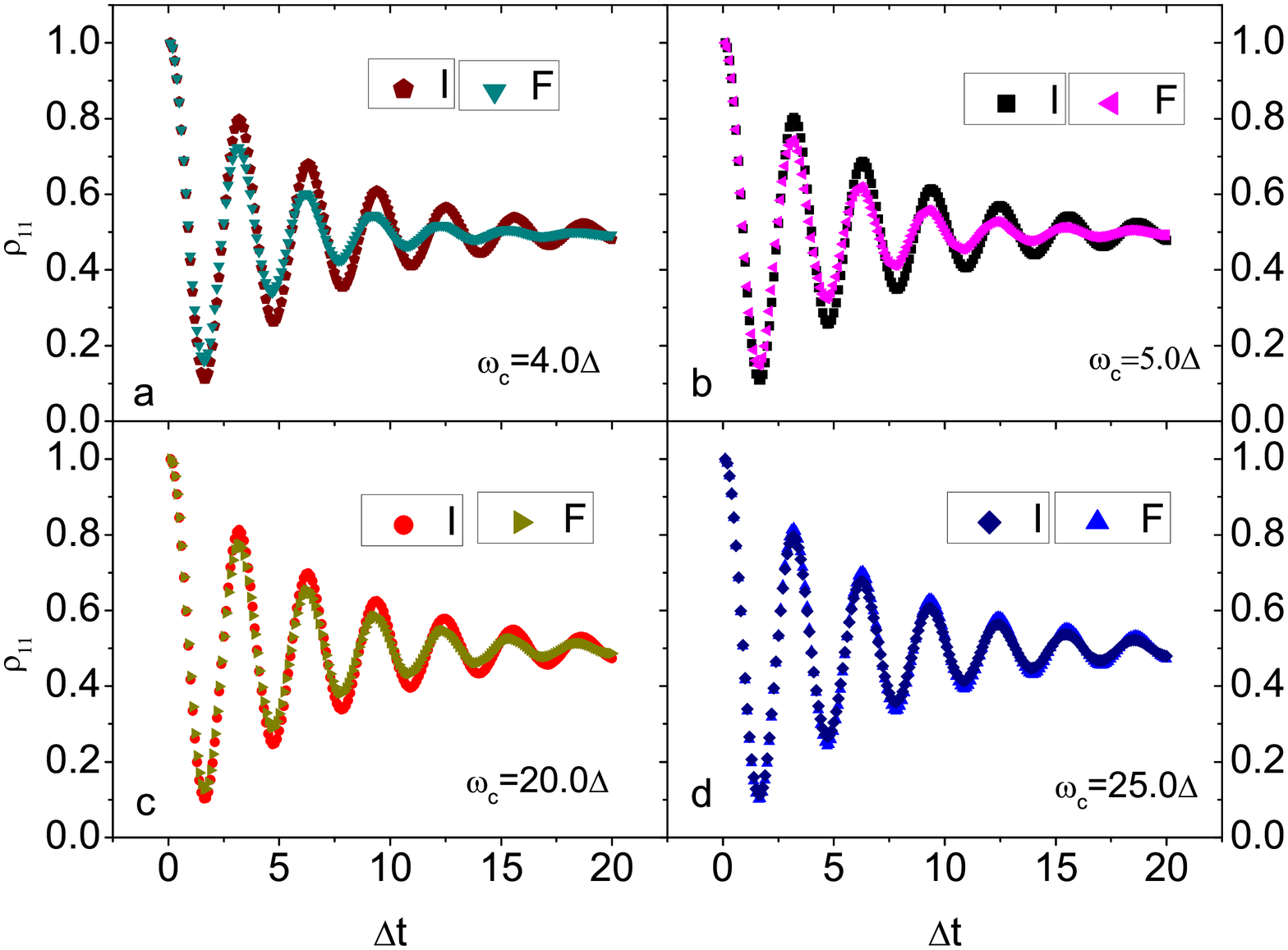}}
\scalebox{0.3}{\includegraphics{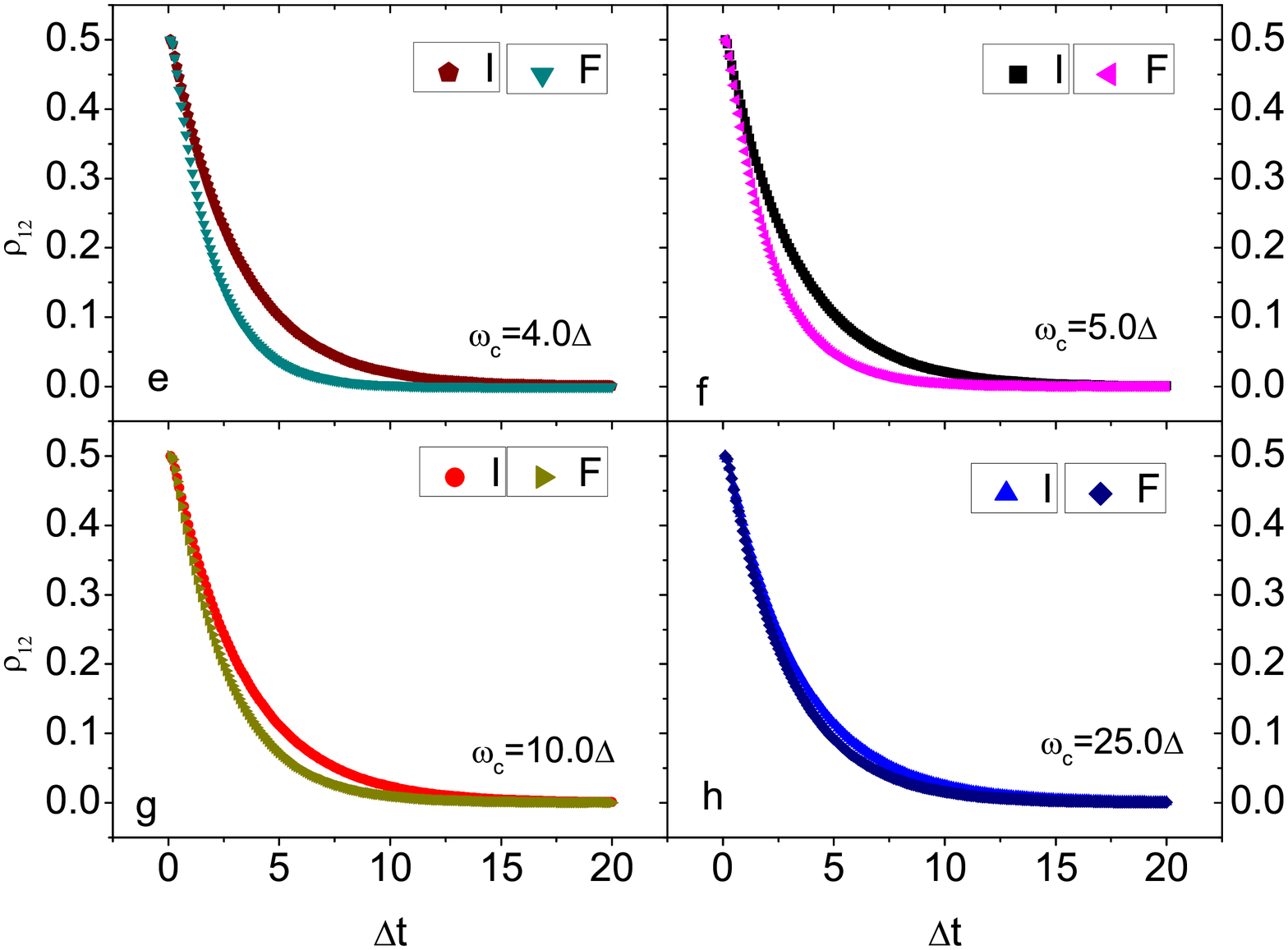}}
\end{center}
\caption{(Color online) The reduced dynamics of qubit in model B
governed by $J_{B}^{I}\left( \omega\right)$ vs by
$J_{B}^{F}\left(\omega\right)$ in different values of
$\omega_{c}$. Here, $\lambda=1, \kappa_{1}=1, \eta'=0.0035,$ and
$\Gamma=52\Delta$, and the values of other parameters are the same
as in Fig. 2.} \label{fig3}
\end{figure}

\begin{figure}[tbp]
\begin{center}
\scalebox{0.4}{\includegraphics{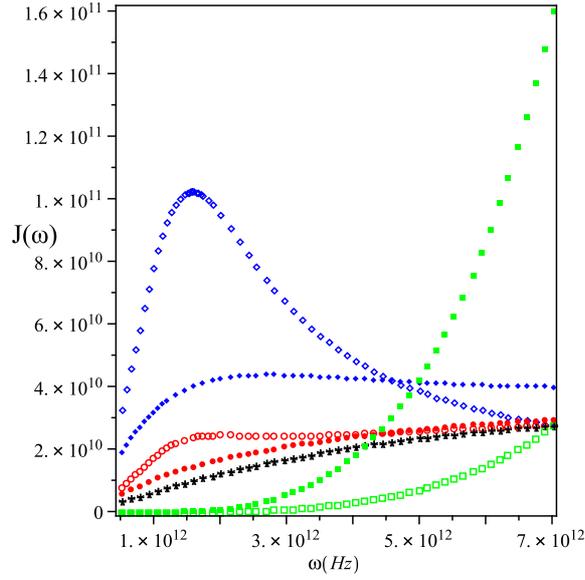}}
\end{center}
\caption{Effective spectral density functions,
$J_{x}^{F}\left(\omega\right)$ in different models $x=A$ (black
asterisk), $x=B$ (blue diamond), $x=C$ (red circle), and $x=D$
(green box), and $J_{x}^{I}\left(\omega\right)$ in different
models $x=A$ (black cross), $x=B$ (blue solid diamond), $x=C$ (red
solid circle), and $x=D$ (green solid box). Here, we set
$\eta=0.02, \lambda=1,\kappa_{1}=\kappa_{2}=1, \Gamma=52\Delta,
\Omega_{0}=10\Delta, \omega_{c}=11\Delta, \omega_{0}=10\Delta$}.
\label{fig5}
\end{figure}

\begin{figure}[tbp]
\begin{center}
\scalebox{0.3}{\includegraphics{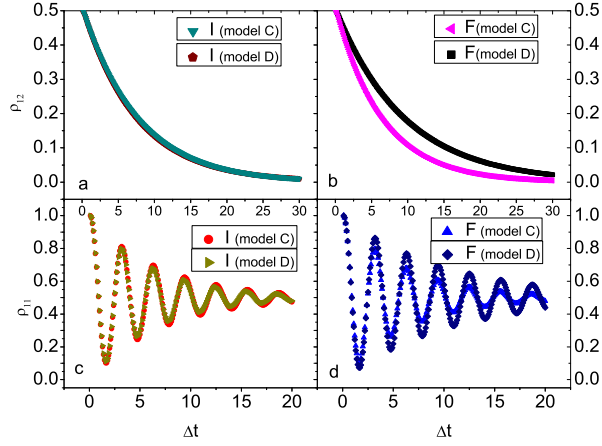}}
\end{center}
\caption{(Color online) The reduced dynamics of the qubit in model
C versus in model D governed by the ESDFs $J_{C}^{I/F}\left(
\omega\right)$ and $J_{D}^{I/F}\left(\omega\right)$. Top two
figures a and b plot the evolutions of the off-diagonal
$\rho_{12}(t)$. Bottom two figures plot the evolutions of the
diagonal elements $\rho_{11}(t)$. Here, $\kappa_{1} =\kappa_{2}
=\lambda=1, \eta'=0.0035$, and $\omega_{c}=7\Delta,
\Omega_{0}=10\Delta,$ and the values of the other parameters are
the same as in Fig 2.} \label{fig4}
\end{figure}

\begin{figure}[tbp]
\begin{center}
\scalebox{0.3}{\includegraphics{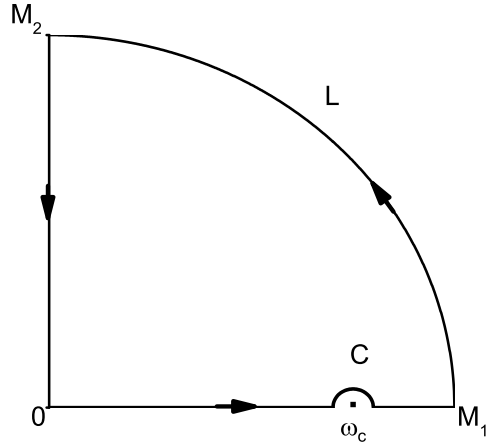}}
\end{center}
\caption{Integral scheme of Eq.\ref{a2}} \label{fig6}
\end{figure}

\end{document}